\documentclass{llncs}

\usepackage[dvips]{graphicx}
\usepackage{amsmath,amssymb,vaucanson-g}

\DeclareMathOperator{\rep}{rep}
\DeclareMathOperator{\val}{val}

\begin{document}

\title{Numeration Systems: a Link between Number Theory and Formal Language Theory}

\author{Michel Rigo}
\institute{Universit\'e de Li\`ege,
Institut de Math\'ematiques,
Grande Traverse 12 (B 37),\\
B-4000 Li\`ege,
Belgium,
\email{M.Rigo@ulg.ac.be}
\thanks{Dedicated to the memory of my grandfather Georges Henderyckx 1930--2010.}}
\maketitle

\begin{abstract}
    We survey facts mostly emerging from the seminal results of Alan
    Cobham obtained in the late sixties and early seventies. We do not
    attempt to be exhaustive but try instead to give some personal
    interpretations and some research directions. We discuss the
    notion of numeration systems, recognizable sets of integers and
    automatic sequences. We briefly sketch some results about
    transcendence related to the representation of real numbers. We
    conclude with some applications to combinatorial game theory and
    verification of infinite-state systems and present a list of open
    problems.
\end{abstract}

\section{Introduction}

It is challenging to give a talk about the interactions existing
between formal language theory and number theory. The topic is vast,
has several entry points and many applications. To cite just a
few: non-adjacent form (NAF) representations to speed up computations arising in elliptic
curve cryptography \cite{NAF}, verification of infinite-state systems \cite{Boigelot&Brusten&Leroux:2009}, combinatorial game theory, fractals and tilings
\cite{PytheasFogg:2002,Berthe&Siegel:2005}, transcendence results, dynamical systems and ergodic theory
\cite[Chap.~5--7]{CANT}, \cite{Barat&Berthe&Liardet&Thuswaldner:2006,Lind&Marcus:1995}. For instance, there
exist tight and fruitful links between properties sought for in
dynamical systems and combinatorial properties of the corresponding
words and languages.

The aim of this paper is to briefly survey some results mostly
emerging from the seminal papers of Cobham of the late
sixties and early seventies \cite{Cobham:1968,Cobham:1969,Cobham:1972}, while also 
 trying to give some
personal interpretations and some research directions. We do not provide an exhaustive survey of the existing literature but we will give some pointers that we hope could be useful to the reader.

When one considers such interactions, the main ingredient is definitely
the notion of numeration system, which provides a bridge between a set
of numbers (integers, real numbers or elements of some other algebraic
structures
\cite{Katai&Szabo:1975,Allouche&Cateland&Gilbert&Peitgen&Shallit&Skordev:1997}) and
formal language theory. On the one hand, arithmetic properties of
numbers or sets of numbers are of interest and on the other hand,
syntactical properties of the corresponding representations may be
studied. To start with, we consider the familiar integer base $k\ge 2$
numeration system.  Any integer $n>0$ is uniquely represented by a
finite word (its $k$-ary representation) $\rep_k(n)=d_\ell\cdots d_0$
over the alphabet $A_k=\{0,\ldots,k-1\}$ such that $\sum_{i=0}^\ell
d_i\, k^i=n$ and $d_\ell\neq 0$. Note that imposing the uniqueness of
the representation allows us to define a map $\rep_k:\mathbb{N}\to
A_k^*$. Nevertheless, in many contexts it is useful to consider all
the representations of an integer $n$ over a given finite alphabet $D\subset\mathbb{Z}$, that is all the words $c_\ell\cdots c_0\in D^*$
such that $\sum_{i=0}^\ell c_i\, k^i=n$. For instance, if $w$ is the
$k$-ary representation of $n$ and if the alphabet $D$ is simply $A_k$, then for all $j>0$, $0^jw$ is another
representation of $n$.

\begin{figure}[htbp]
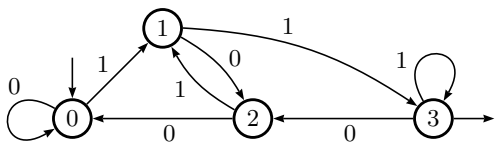

        \centering
\VCDraw{%
        \begin{VCPicture}{(0,-0.5)(8,2)}
 \State[0]{(0,0)}{0}  \State[1]{(2,2)}{1}
 \State[2]{(4,0)}{2} \State[3]{(8,0)}{3}
\Initial[n]{0}
\Final[e]{3}
\LoopW{0}{0}
\EdgeL{0}{1}{1}
\ArcL[.6]{1}{2}{0}
\ArcL{1}{3}{1}
\EdgeL{2}{0}{0}
\ArcL[.6]{2}{1}{1}
\EdgeL{3}{2}{0}
\LoopN{3}{1}
\end{VCPicture}
}
\caption{A finite automaton accepting $0^*\rep_2(4\mathbb{N}+3)$.}
        \label{chcob:fig:1}
    \end{figure}

\begin{figure}[htbp]
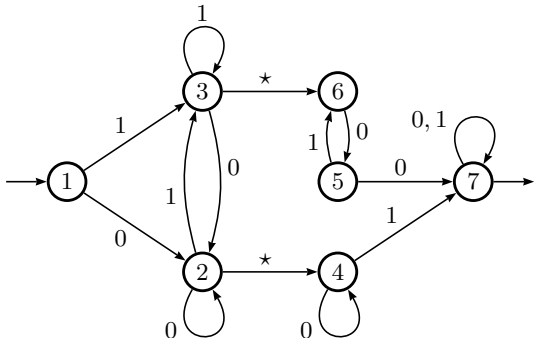

        \centering
\VCDraw{%
        \begin{VCPicture}{(0,-3)(9,2.7)}
\State[1]{(0,0)}{1}  
\State[3]{(3,2)}{3}
\State[2]{(3,-2)}{2} 
\State[4]{(6,-2)}{4}
\State[5]{(6,0)}{5}
\State[6]{(6,2)}{6}
\State[7]{(9,0)}{7}
\Initial[w]{1}
\Final[e]{7}
\EdgeR{1}{2}{0}
\EdgeL{1}{3}{1}
\LoopS{2}{0}
\LoopN[.5]{3}{1}
\EdgeL{2}{4}{\star}
\EdgeL{3}{6}{\star}
\ArcL{2}{3}{1}
\ArcL{3}{2}{0}
\ArcL{5}{6}{1}
\ArcL{6}{5}{0}
\LoopS{4}{0}
\EdgeL{4}{7}{1}
\EdgeL{5}{7}{0}
\LoopN{7}{0,1}
\end{VCPicture}
}
\caption{A B\"uchi automaton accepting   
$\{2n+(0,4/3)\mid n\in\mathbb{Z}\}$.}
        \label{fig:1nn}
    \end{figure}

In the same way, any real number $r\in(0,1)$ is represented by an
infinite word $d_1d_2\cdots$ over $A_k$ such that
$\sum_{i=1}^{+\infty} d_i\, k^{-i}=r$. Uniqueness of the
representation may be obtained by taking the maximal word for the
lexicographic ordering on $A_k^\omega$ satisfying the latter equality;
in this case, the sequence $(d_i)_{i\ge 1}$ is not ultimately constant
and equal to $k-1$: there is no $N$ such that, for all $n\ge N$, $d_n=k-1$. Therefore, to represent a real number $r>0$, we
take separately its integer part $\lfloor r\rfloor$ and its fractional
part $\{r\}$.  Base $k$-complements or signed number representations
\cite{Knuth:1981} can be used to represent negative elements as well,
the sign being determined by the most significant digit which is thus
$0$ or $k-1$.  By convention, the empty word $\varepsilon$ represents
$0$, i.e., $\rep_k(0)=\varepsilon$. If the numeration system is fixed,
say the base $k$ is given, then any integer $n$ (resp. any real number
$r>0$) corresponds to a finite (resp. infinite) word over $A_k$ (resp.
over $A_k\cup\{\star\}$, where $\star$ is a new symbol used as a
separator). Therefore any set of numbers corresponds to a language of
representations and we naturally seek for some link between the
arithmetic properties of the numbers belonging to the set and the
syntactical properties of the corresponding representations. Let $X$
be a subset of $\mathbb{N}$. Having in mind Chomsky's hierarchy, the
set $X$ could be considered quite ``simple'' from an algorithmic point
of view whenever the set of $k$-ary representations of the elements in
$X$ is a regular (or rational) language accepted by a finite
automaton. A set $X\subseteq\mathbb{N}$ satisfying this property is
said to be {\em $k$-recognizable}. Note that $X$ is $k$-recognizable
if and only if $0^*\rep_k(X)$ is regular. As an example, a DFA (i.e.,
a deterministic finite automaton) accepting exactly the binary
representations of the integers congruent to $3$ (mod~$4$) is given in
Figure~\ref{chcob:fig:1}.
    Similarly, a set $X\subseteq\mathbb{R}$ of real numbers is
    {\em $k$-recognizable} if there exists a finite (non-deterministic) B\"uchi automaton accepting {\em all} the
    $k$-representations over $A_k$ of the elements in $X$, that is, the representations starting with an arbitrary number of leading zeroes, and in particular the ones ending with
    $(k-1)^\omega$. Such an automaton is often
    called a {\em Real Number Automaton}
    \cite{Boigelot&Rassart&Wolper:1998}. The B\"uchi automaton in
    Figure~\ref{fig:1nn} (borrowed from a talk given by
    V.~Bruy{\`e}re) accepts all the possible binary encodings (using
    base $2$-complement for negative numbers) of elements in the set
    $\{2n+(0,4/3)\mid n\in\mathbb{Z}\}$. For instance $3/2$ is encoded
    by the language of infinite words $0^+1\star 10^\omega\cup
    0^+1\star 01^\omega$. Note that the transition
    $3\stackrel{\star}{\longrightarrow}6$ (resp. $2\stackrel{\star}{\longrightarrow}4$) is considered for an odd (resp. even)
    integer part and the series $\sum_{i=1}^{+\infty} 4^{-i}=1/3$ corresponds to the cycle $\{5,6\}$.

    To generalize the $k$-ary integer base system, it is quite natural
    to consider an increasing sequence of integers, like the Fibonacci
    sequence, as a numeration basis to get a greedy decomposition of
    any integer (see Definition~\ref{def:numsys}) or the negative
    powers of a real number $\beta>1$ to develop any real $r\in(0,1)$
    as $\sum_{i=1}^{+\infty} c_i\, \beta^{-i}$ with the coefficients
    $c_i$ belonging to a convenient finite alphabet. Let us point out
    Fraenkel's paper \cite{Fraenkel:1985} for some general ideas about
    representations of integers in various numeration systems.
    Among non-standard decompositions of integers, let us mention
    the so-called {\em combinatorial numeration system} going back to
    Lehmer and Katona, where integers are decomposed using binomial
    coefficients with some prescribed property, also see
    \cite{Charlier&Rigo&Steiner:2008}, and the {\em factorial
      numeration system} \cite{Lenstra:2005}.  In Frougny and
    Sakarovitch's chapter \cite[Chap.~2]{CANT} and in Frougny's chapter 
    \cite[Chap.~7]{Lothaire:2002} many details on recognizable sets
    and about the representation of integers and real numbers are
    given. In particular, 
    Parry's $\beta$-developments of real numbers \cite{Parry:1960}  are presented in the latter reference.
    Abstract numeration systems (see Definition~\ref{def:abstract})
    are a general framework to study recognizable sets of integers,
    see \cite{Lecomte&Rigo:2001} and \cite[Chap.~3]{CANT}.

\medskip

The seminal work of Cobham may be considered as a starting point for
the study of recognizable sets for at least three reasons. Let us
sketch these below. Details and definitions will be given in the next
sections.
\begin{itemize}
  \item[\textup{(i)}] Cobham's theorem from 1969 \cite{Cobham:1969}
    states that the recognizability of a set of integers, as
    introduced above, strongly depends on the choice of the base, e.g., there
    are sets which are $2$-recognizable but not $3$-recognizable. The
    only subsets of $\mathbb{N}$ that are recognizable in all bases
    are exactly the ultimately periodic sets, i.e., the finite unions
    of arithmetic progressions. See Theorem~\ref{the:cobham1} in
    Section~\ref{sec:cob1} below for the exact statement of the
    result. It is an easy exercise to show that an arithmetic
    progression is $k$-recognizable for all $k\ge 2$ (e.g., Figure~\ref{chcob:fig:1}). See for instance
    \cite[prologue]{Sakarovitch:2003} about the {\em machine {\`a} diviser de Blaise Pascal}. In that direction, an
    interesting question \cite{Alexeev:2004} is to obtain the
    state complexity of the minimal automaton recognizing a given
    divisibility criterion in an integer base. For this state
    complexity question studied in the wider context of linear
    numeration systems (cf. Definition~\ref{def:linear}), see
    \cite{Charlier&Rampersad&Rigo&Waxweiler}.

    The base dependence of recognizability shown by Cobham's result
    strongly motivates the general study of recognizable sets and the
    introduction of non-standard or exotic numeration systems based on
    an increasing sequence satisfying a linear recurrence relation.

    For integer base $k$ numeration systems, nice characterizations of
    recognizable sets are well-known: logical characterization by
    first order formulas in a suitable extension of the Presburger
    arithmetic $\langle\mathbb{N},+\rangle$, $k$-automatic
    characteristic sequence generated through a uniform morphism of length $k$, characterization in terms of algebraic formal power series for a prime
    base. See the excellent survey
    \cite{Bruyere&Hansel&Michaux&Villemaire:1994} where the so-called Cobham--Semenov' theorem, which extends
     Cobham's original result from 1969 to subsets of $\mathbb{N}^d$,
    $d\ge 2$,  is also
    presented. Recall that the {\em characteristic sequence}
    $(x_i)_{i\ge 0}\in\{0,1\}^{\mathbb{N}}$ of $X\subseteq\mathbb{N}$ is defined by $x_i=1$
    if and only if $i\in X$. It is not our goal to give here a full
    list of pointers to the existing bibliography on the ramifications
    of Cobham's theorem, see for instance \cite{Durand&Rigo:2010}. For
    presentations of Cobham's theorem based on Georges Hansel's work,
    see \cite{Perrin:1990,Allouche&Shallit:2003} together with
    \cite{Rigo&Waxweiler:2006}.

\smallskip

  \item[\textup{(ii)}] The next paper of Cobham from 1972
    \cite{Cobham:1972} introduced the concept of $k$-automatic
    sequences (originally called tag sequences, see
    Definition~\ref{def:sub:autom}) and linked numeration systems with
    the so-called substitutions and morphic words (see
    Definition~\ref{def:sub:morphic}).  It is easy to see that a set
    $X\subseteq\mathbb{N}$ is $k$-recognizable if and only if the
    characteristic sequence of $X$ is a $k$-automatic infinite word
    over $\{0,1\}$. For a comprehensive book on $k$-automatic
    sequences, see \cite{Allouche&Shallit:2003}. As we will see, this
    approach gives another way to generalize the notion of
    a recognizable set by considering sets having a morphic
    characteristic sequence (see Remark~\ref{rem:mo}).  Details will be presented in
    Section~\ref{sec:sub}.

\smallskip

\item[\textup{(iii)}] As the reader may already have noticed, this
  survey is mainly oriented towards sets of numbers (integers) giving
  rise to a language of representations. Another direction should be
  to consider a single real number and the infinite word representing
  it in a given base. Alan Cobham also conjectured the following
  result proved later on by Adamczewski and Bugeaud. {\em Let $\alpha$
    be an algebraic irrational real number. Then the base-$k$
    expansion of $\alpha$ cannot be generated by a finite automaton.}
  Cobham's question follows a question of Hartmanis and Stearns \cite{hartmanis}: {\em
    does it exist an algebraic irrational number computable in linear
    time by a (multi-tape) Turing machine?} In the same
    way, if an infinite word $w$ over the finite alphabet $A_k$ of
    digits has some specific combinatorial properties (like, a low factor
    complexity, or being morphic or substitutive), is
    the corresponding real number having $w$ as $k$-ary representation
    transcendental? Let us mention that several surveys in that
    direction are worth of reading \cite[Chap.~10]{Lothaire:2005},
    \cite[Chap.~8]{CANT},
    \cite{Adamczewski&Bell:2010,Waldschmidt:2009}.  We will briefly
    sketch some of these developments in Section~\ref{sec:trans}.
\end{itemize}

In Section~\ref{sec:4}, we sketch some of the links existing between
numeration systems, combinatorics on words and combinatorial game
theory. Cobham's theorem about base dependence also appears in the
framework of the verification of infinite state systems, see
Section~\ref{sec:5}.  Finally, in Section~\ref{sec:6} we give some
paths to the literature that the interested reader may follow and in
Section~\ref{sec:7}, we present some open questions.

\section{Cobham's Theorem and Base Dependence}\label{sec:cob1}

Two integers $k,\ell\ge 2$ are {\em multiplicatively independent} if
the only integers $m,n$ such that $k^m=\ell^n$ are $m=n=0$. Otherwise
stated, $k,\ell\ge 2$ are multiplicatively independent if and only if
$\log k/\log \ell$ is irrational. Recall a classical result in
elementary number theory, known as Kronecker's theorem: {\em if $\theta>0$
is irrational, then the set $\{ \{n\theta\} \mid n>0\}$ is dense in
$[0,1]$}. Such result is an important argument in the proof of the following result.

\begin{theorem}[Cobham's theorem \cite{Cobham:1969}]\label{the:cobham1}
    Let $k,\ell\ge 2$ be two multiplicatively independent integers. A set
    $X\subseteq\mathbb{N}$ is simultaneously $k$-recognizable and
    $\ell$-recognizable if and only if $X$ is ultimately periodic.
\end{theorem}

Obviously the set $P_2=\{2^n\mid n\ge 1\}$ of powers of two is
$2$-recognizable because $\rep_2(P_2)=10^*$. But since $P_2$ is not
ultimately periodic, Cobham's theorem implies that $P_2$ cannot be
$3$-recognizable. To see that a given infinite ordered set
$X=\{x_0<x_1<x_2<\cdots\}$ is $k$-recognizable for {\em no} base $k\ge
2$ at all, we can use results like the following one where the
behavior of the ratio ({\em resp.}  difference) of any two consecutive
elements in $X$ is studied through the quantities
$$\mathbf{R}_X:=\limsup_{i \rightarrow \infty}\, \frac{x_{i+1}}{x_i}\text{ and }
\mathbf{D}_X:=\limsup_{i \rightarrow \infty}\, (x_{i+1}-x_i).$$

\begin{theorem}[Gap theorem \cite{Cobham:1972}]
    Let $k\ge 2$. If $X\subseteq\mathbb{N}$ is a $k$-recognizable infinite subset of
    $\mathbb{N}$, then either $\mathbf{R}_X>1$ or $\mathbf{D}_X<+\infty$.
\end{theorem}

\begin{corollary}\label{chcob:cor:gap} Let $a\in\mathbb{N}_{\ge 2}$.  The set of primes and
    the set $\{n^a\mid n\ge 0\}$ are never $k$-recognizable for any
    integer base $k\ge 2$.
\end{corollary}

Proofs of the Gap theorem and its corollary can also be found in
\cite{Eilenberg:1974}. It is easy to show that $X\subseteq\mathbb{N}$ is $k$-recognizable if and only if it is $k^n$-recognizable, $n\in\mathbb{N}\setminus\{0\}$. As a  consequence of Cobham's theorem, sets
of integers can be classified into three categories:
\begin{itemize}
  \item ultimately periodic sets which are recognizable for all bases,
  \item sets which are $k$-recognizable for some $k\ge 2$, and which
    are $\ell$-recognizable only for those $\ell\ge 2$ such that $k$ and
    $\ell$ are multiplicatively dependent bases, for example, the set $P_2$ of powers of two, 
  \item sets which are $k$-recognizable for no base $k\ge 2$ at all, for example, the set of squares.
\end{itemize}

\begin{definition}
    An infinite ordered set $X=\{x_0<x_1<x_2<\cdots\}$ such that
    $\mathbf{D}_X<+\infty$ is said to be {\em syndetic} or with {\em
      bounded gaps}. Otherwise stated, $X$ is syndetic if there exists
    $C>0$ such that, for all $n\ge 0$, $x_{n+1}-x_n<C$.
\end{definition}

If $X\subseteq\mathbb{N}$ is ultimately periodic, then $X$ is
syndetic. Note that the converse does not hold. For instance, consider
the complement of the set $\{2^n\mid n\ge 0\}$ which is syndetic,
$2$-recognizable but not ultimately periodic.
\begin{example}[Thue--Morse set]\label{exa:tm} Let $n\in\mathbb{N}$.  Denote by
    $s_k(n)$ the classical number-theoretic function summing up the
    digits appearing in $\rep_k(n)$. As a classical example, consider
    the set $T=\{n\in\mathbb{N}\mid s_2(n)\equiv 0\bmod{2}\}$. This
    set is $2$-recognizable and syndetic but not ultimately periodic. 
    It appears in several contexts \cite{Allouche&Shallit:1999} and in
    particular, it provides a solution to Prouhet's problem (also
    known as the Prouhet--Tarry--Escott problem which is a special case of
    a multi-grade equation). 
 
\end{example}

The set of squares is $k$-recognizable for no integer base $k$ but as
we shall see this set is recognizable for some non-standard numeration
systems (see Example~\ref{exa:squares}). One possible extension of $k$-ary numeration systems is to consider a
numeration basis.

\begin{definition}\label{def:numsys}
    A {\em numeration basis} is an increasing sequence $U=(U_n)_{n\ge
      0}$ of integers such that $U_0=1$ and $\sup_{i\ge 0} U_{i+1}/U_i$ is
    bounded. 
\end{definition}
Using the greedy algorithm, any integer $n>0$ has a unique decomposition 
$$n=\sum_{i=0}^\ell c_i\, U_i$$
where the coefficients $c_i$ belong to the finite set
$A_U=\{0,\ldots, \sup \lceil U_{i+1}/U_i\rceil-1\}$. Indeed there exists a
unique $\ell\ge 0$ such that $U_{\ell}\le n< U_{\ell+1}$. Set
$r_\ell=n$. For all $i=\ell,\ldots,1$, proceed to the Euclidean
division $r_i = c_i\, U_i + r_{i-1}$, with $r_{i-1}<U_i$. The word
$c_\ell\cdots c_0$ is the (normal) {\em $U$-representation} of $n$ and is
denoted by $\rep_U(n)$.  Naturally, these non-standard numeration
systems include the usual integer base $k$ system by taking
$U_n=k^n$ for all $n\ge 0$. The {\em numerical value map} $\val_U:A_U^*\to\mathbb{N}$ maps any
word $d_\ell\cdots d_0$ over $A_U$ onto $\sum_{i=0}^\ell d_iU_i$.

\begin{remark}
    By contrast with abstract numeration systems that will be
    introduced later on, when dealing with a numeration basis we
    often use the terminology of a {\em positional numeration system}
    to emphasize the fact that a digit $d\in A_U$ in the $i$th position
    (counting from the right, i.e., considering the least significant digit first)
    of a $U$-representation has a weight $d$ multiplied by the
    corresponding element $U_i$ of the basis.
\end{remark}

Having in mind the notion of $k$-recognizable
sets, a set $X\subseteq\mathbb{N}$ is said to be {\em $U$-recognizable} if $\rep_U(X)=\{\rep_U(n)\mid n\in X\}$ is a regular language over the alphabet $A_U$. Note that $\rep_U(X)$ is regular if and only if $0^*\rep_U(X)$ is regular.

\begin{definition}\label{def:linear}
    A numeration basis $U=(U_n)_{n\ge 0}$ is said to be {\em linear}, 
    if there exist $a_0,\ldots,a_{k-1}\in\mathbb{Z}$ such that
\begin{equation}
    \label{eq:rec}
    \forall n\ge 0,\ U_{n+k}=a_{k-1}U_{n+k-1}+\cdots +a_0 U_n.
\end{equation}
If
$\lim_{n\to\infty}U_{n+1}/U_n=\beta$ for some real $\beta> 1$, then $U$ is
said to {\em satisfy the dominant root condition} and $\beta$ is called the
{\em dominant root} of the recurrence.
\end{definition}

If $\mathbb{N}$ is $U$-recognizable, then $U$ is a linear numeration
basis \cite{Shallit:1994,CANT} (hint: observe that $\rep_U(\{U_n\mid n\ge 0\})=10^*$). For a discussion on sufficient
conditions on the recurrence relation satisfied by $U$ for the
$U$-recognizability of $\mathbb{N}$, see \cite{Hollander:1998} and \cite{Loraud:1995}. In
particular, as shown by the next result, {\em for a linear numeration basis $U$, the set $\mathbb{N}$ is
$U$-recognizable if and only if all ultimately periodic sets are
$U$-recognizable}.

\begin{theorem}[Folklore \cite{CANT}]\label{the:folk}
    Let $p,r\ge 0$. If $U=(U_n)_{n\ge 0}$ is a linear numeration
    basis, then
$$\val_U^{-1}(p\,\mathbb{N}+r)=\biggl\{c_\ell\cdots c_0\in A_U^*\mid 
\sum_{k=0}^\ell c_k\, U_k\in p\, \mathbb{N}+r\biggr\}$$
is accepted by a DFA that can be effectively constructed. In
particular, if $\mathbb{N}$ is $U$-recognizable, then any ultimately periodic
set is $U$-recognizable.
\end{theorem}

\begin{example}\label{exa:ns-fib}
    Consider the Fibonacci numeration system given by the basis
    $F_0=1$, $F_1=2$ and $F_{n+2}=F_{n+1}+F_n$ for all $n\ge 0$. For this system,
    $0^*\rep_F(\mathbb{N})$ is given by the set of words over
    $\{0,1\}$ avoiding the factor $11$ and  the set of even numbers is
    $U$-recognizable \cite{Charlier&Rampersad&Rigo&Waxweiler} using the DFA shown in
    Figure~\ref{fig:fib2N}.
\begin{figure}[htbp]
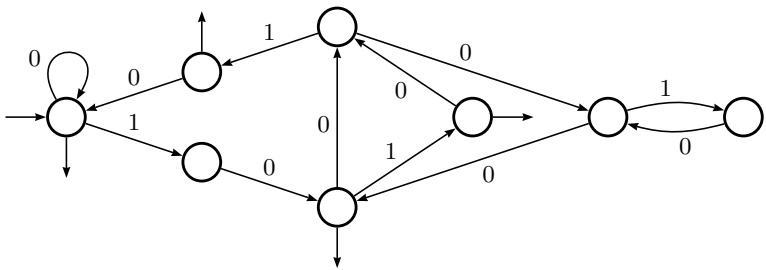

        \centering
\VCDraw{%
        \begin{VCPicture}{(0,-3)(15,2)}
 \State[]{(0,0)}{1} 
 \State[]{(3,-1)}{2} 
 \State[]{(6,-2)}{4} 
 \State[]{(6,2)}{5} 
 \State[]{(9,0)}{6} 
 \State[]{(12,0)}{7} 
 \State[]{(3,1)}{8}
 \State[]{(15,0)}{9}  
\Initial[w]{1}
\Final[s]{1}
\Final[s]{4}
\Final[e]{6}
\Final[n]{8}

\LoopN{1}{0}
\EdgeL{1}{2}{1}
\EdgeL{2}{4}{0} 
\EdgeL{4}{5}{0} 
\EdgeL{4}{6}{1}  
\EdgeL{5}{7}{0} 
\EdgeR{5}{8}{1}  
\EdgeL{6}{5}{0}
\EdgeL{7}{4}{0} 
\ArcL{7}{9}{1} 
\EdgeR{8}{1}{0} 
\ArcL{9}{7}{0} 
\end{VCPicture}
}
\caption{A finite automaton accepting $0^*\rep_F(2\mathbb{N})$.}
        \label{fig:fib2N}
    \end{figure}
\end{example}

To conclude this section, we present a linear numeration basis $U$
such that the set of squares $\mathcal{Q}=\{n^2\mid n\in\mathbb{N}\}$ is
$U$-recognizable. This set will also be used in
Example~\ref{exa:squares} to get a set having a morphic characteristic
sequence.

\begin{example}
    Consider the sequence given $U_n=(n+1)^2$ for all $n\ge 0$. This
    sequence satisfies, for all $n\ge 0$, the relation $U_{n+3}=3\,
    U_{n+2}-3\, U_{n+1}+U_n$.  In that case, $\rep_U(\mathbb{N})\cap
    10^*10^*=\{10^a10^b\mid b^2<2a+4\}$ showing with the pumping lemma
    that $\mathbb{N}$ is not $U$-recognizable \cite{Shallit:1994}. But
    trivially, we have $\rep_U(\mathcal{Q})=10^*$.
\end{example}

\section{Substitutions and Abstract Numeration Systems}\label{sec:sub}

For basic facts on morphisms over $A^*$ or the usual distance put on
$A^\omega$ (which gives a notion of convergence), see classic textbooks like
\cite{Allouche&Shallit:2003,CANT,Lothaire:2002}. Let $A$ be a finite alphabet and $\sigma:A^*\to A^*$ be a
    morphism.  If there exist a letter $a\in A$ and a word $u\in A^+$
    such that $\sigma(a)=au$ and moreover, if
    $\lim_{n\to+\infty}|\sigma^n(a)|=+\infty$, then $\sigma$ is said
    to be {\em prolongable} on $a$. 

\begin{definition}\label{def:sub:morphic}
     Let $\sigma:A^*\to A^*$ be a
    morphism prolongable on $a$.  We have
     $$\sigma(a)=a\, u,\ \sigma^2(a)=a\, u\, \sigma(u),\ \sigma^3(a)=a\, u\, \sigma(u)\, 
     \sigma^2(u),\ \ldots\  .    $$ Since, for all $n\in\mathbb{N}$, $\sigma^n(a)$ is a prefix of $\sigma^{n+1}(a)$
     and because $|\sigma^n(a)|$ tends to infinity when $n\to+\infty$, the
    sequence $(\sigma^n(a))_{n\ge 0}$ converges to an infinite word
    denoted by $\sigma^\omega(a)$ and given by
    $$\sigma^\omega(a):=\lim_{n\to+\infty}\sigma^n(a)=a\, u\,
    \sigma(u)\, \sigma^2(u)\, \sigma^3(u)\cdots \ . $$ This infinite
    word is a fixed point of $\sigma$.  An infinite word obtained in
    this way by iterating a prolongable morphism is said to be {\em
      purely morphic}.  In the literature, one also finds the term
    {\em pure morphic}.  If $x\in A^\mathbb{N}$ is purely morphic and
    if $\tau:A\to B$ is a coding (or letter-to-letter morphism), then
    the word $y=\tau(x)$ is said to be {\em morphic}.
  \end{definition} 

  \begin{definition}\label{def:sub:autom}
  Let $k\ge 2$. A morphic word $w\in B^\omega$ is {\em $k$-automatic}
  if there exists a morphism $\sigma:A^*\to A^*$ and a coding $\tau$
  such that $w=\tau(\sigma^\omega(a))$ and, for all $c\in A$,
  $|\sigma(c)|=k$. A morphism satisfying this latter property is said to be {\em uniform}. \end{definition}

 The link between $k$-recognizable sets and
  $k$-automatic sequences is given by the following result. In particular, in the proof of this result, it is interesting to note that an automaton is canonically associated with a morphism.

  \begin{theorem}\cite{Cobham:1972}\label{the:cob2}
      An infinite word $w=w_0w_1w_2\cdots$ over an alphabet $A$ is
      $k$-automatic if and only if, for all $a\in A$, the set
      $X_a=\{i\in\mathbb{N}\mid w_i=a\}$ is $k$-recognizable.
  \end{theorem}

  Otherwise stated, $w=w_0w_1w_2\cdots\in A^\omega$ is $k$-automatic if and only
  if there exists a deterministic finite automaton with output (DFAO)
  $\mathcal{M}$ where $Q$ is the set of states of $\mathcal{M}$, $\delta:Q\times A_k\to Q$ (resp. $\tau:Q\to A$) is the
  transition function (resp. output function) of $\mathcal{M}$, such that $\tau(\delta(q_0,\rep_k(n))=w_n$ for all
  $n\ge 0$.

\begin{remark}\label{rem:mo}
    Using automata as a model of computation, $U$-recognizable sets
    naturally raise some interest. On the same level, sets of integers
    having a morphic characteristic sequence can be considered as
    another natural generalization of the concept of
    $k$-recognizability.  Iterations of a morphism may be used to get
    inductively further elements of the set defined by the morphism
    and a coding. As will be shown by Theorem~\ref{the:cob3}, similarly to the case of uniform morphisms (as given in Definition~\ref{def:sub:autom}) described above, the
    computation of a given element can also be done by using a DFAO
    and representations of integers in an abstract numeration
    system.
\end{remark}
    
\begin{example}\label{exa:squares}
    Consider the alphabet $A=\{a,b,c\}$ and the morphism
    $\sigma:A^*\to A^*$ defined by $\sigma:a\mapsto abcc,b\mapsto
    bcc,c\mapsto c$. We get
$$\sigma^\omega(a)=abccbccccbccccccbccccccccbccccccccccbcc\cdots\ .$$
It is easy to see that considering the coding $\tau:a,b\mapsto 1$ and $\tau:c\mapsto 0$, the word $\tau(\sigma^\omega(a))$ is the characteristic sequence of the set of squares.
\end{example}

The {\em factor complexity} of an infinite word $w$ is the non-decreasing function\linebreak
$p_w:\mathbb{N}\to\mathbb{N}$ mapping $n$ onto the number of distinct
factors (or subwords) occurring in $w$. See for instance
\cite[Chap.~4]{CANT}. For a survey on the factor complexity of morphic
words, see \cite{Allouche:1994}. In 1972, Cobham already observed that
if $w$ is $k$-automatic, then $p_w$ is in $\mathcal{O}(n)$. For
instance, the factor complexity of the characteristic sequence of the Thue--Morse set $T$ considered in
Example~\ref{exa:tm} is computed in
\cite{Brlek:1989,deLuca&Varricchio:1988}.

\begin{theorem}[Morse--Hedlund's Theorem]
 Let $x=x_0x_1x_2\dotsm$ be an infinite word over~$A$.
 The following conditions are equivalent.
 \begin{itemize}
   \item The complexity function~$p_x$ is bounded by a constant,
     i.e., there exists $C$ such that for all $n\in\mathbb{N}$, $p_x(n)\leq C$.
   \item There exists $N_0\in\mathbb{N}$ such that for all $n\geq N_0$,
     $p_x(n)=p_x(N_0)$.
   \item There exists $N_0\in\mathbb{N}$ such that $p_x(N_0)=N_0$.
   \item There exists $m\in\mathbb{N}$ such that $p_x(m)=p_x(m+1)$.
   \item The word~$x$ is ultimately periodic.
 \end{itemize}    
\end{theorem}
In particular, non ultimately periodic sequences with low complexity
are the so-called {\it Sturmian sequences} whose factor complexity is
$p(n)=n+1$ for all $n\ge 1$. Note that such sequences are over a
binary alphabet, $p(1)=2$. For a survey on Sturmian words, see for
instance \cite{Lothaire:2002}. Since Pansiot's work \cite{Pansiot:1984}, the factor
complexity of a non ultimately periodic purely morphic word $w$ is well-known, see for instance
\cite[Chap.~4]{CANT} or the survey \cite{Allouche:1994}, there exists
constants $C_1,C_2$ such that $C_1f(n)\le p_w(n)\le C_2 f(n)$ where
$f(n)\in\{n,n\log n,n\log\log n,n^2\}$.

\begin{remark}\label{rem:coo}
    F.~Durand has achieved a lot of work towards a general version of
    Cobham's theorem for morphic words
    \cite{Durand:1998b,Durand:1998c,Durand:2002}. Without giving much
    details (see for instance \cite{Durand&Rigo:2010} for a detailed
    account), with a non-erasing morphism $\sigma$ over
    $A=\{a_1,\ldots,a_t\}$ (i.e., $\sigma(\sigma_i)\neq\varepsilon$
    for all $i$) generating a morphic word $w$ (also using an extra
    coding) is associated a matrix $\mathbf{M}_\sigma$ (like the
    adjacency matrix of a graph) where, for all $i,j$,
    $(\mathbf{M}_\sigma)_{i,j}$ is the number of occurrences of the
    letter $a_i$ in the image $\sigma(a_j)$. Considering the morphism
    in Example~\ref{exa:squares}, we get
$$\mathbf{M}_\sigma=
\begin{pmatrix}
    1&0&0\\
    1&1&0\\
    2&2&1\\
\end{pmatrix}.$$
Then considering the irreducible components (i.e., the strongly
connected components of the associated automaton) of the matrix
$\mathbf{M}_\sigma$ and the theorem of Perron--Frobenius, a real
number $\beta>0$ is associated with the morphism. The word $w$ is
therefore said to be {\em $\beta$-substitutive}.  Let $\alpha,\beta>1$
be two multiplicatively independent Perron numbers (the notion of
multiplicative independence extends to real numbers $>1$).  Under some
mild assumptions \cite{Durand&Rigo:2010}, {\em if $w$ is both
  $\alpha$-substitutive and $\beta$-substitutive, then it is
  ultimately periodic}. It is a natural generalization of the fact
that if $k,\ell\ge 2$ are multiplicatively independent, then a word
which is both $k$-automatic and $\ell$-automatic is ultimately
periodic.
\end{remark}

\begin{example}
    The consecrated Fibonacci word, i.e., the unique fixed point of
    $\sigma:0\mapsto 01,1\mapsto 0$, is $\alpha$-substitutive where
    $\alpha$ is the Golden ratio $(1+\sqrt{5})/2$. Therefore, this
    infinite word is $k$-automatic for no integer $k\ge 2$. Indeed,
    $k$ and the Golden ratio are multiplicatively independent.
\end{example}

In view of Theorem~\ref{the:folk}, it is desirable for a numeration
basis $U$ that the set $\mathbb{N}$ be $U$-recognizable. In that case, one can use a finite automaton to test whether or not a given word over $A_U$ is a valid $U$-representation. Taking this requirement as a
basic assumption and observing that for all integers $x,y$, we have
$x<y$ if and only if $\rep_U(x)$ is genealogically less than
$\rep_U(y)$, we introduce the concept of an abstract numeration system.
To define the {\em genealogical order} (also called {\em radix} or {\em military order}),
first order words by increasing length and for words of the same
length, take the usual lexicographical order induced by the ordering of
the alphabet.

\begin{definition}\label{def:abstract}
    Let $L$ be an infinite regular language over a totally ordered
    alphabet $(A,<)$. An {\em abstract numeration system} is the
    triple $S=(L,A,<)$. Ordering by increasing genealogical order the
    words in $L$ provides a one-to-one correspondence between $L$ and
    $\mathbb{N}$. The $n$th word in $L$ (starting from $0$) is denoted
    by $\rep_S(n)$ and the inverse map $\val_S:L\to\mathbb{N}$ is such
    that $\val_S(\rep_S(n))=n$. Any numeration basis $U$ such that
    $\mathbb{N}$ is $U$-recognizable is a particular case of an abstract
    numeration system.  In this respect, a set
    $X\subseteq\mathbb{N}$ is {\em $S$-recognizable}, if $\rep_S(X)$
    is a regular language.

    A sequence $w=w_0w_1\cdots$ is {\em $S$-automatic} if there exists a
    DFAO $\mathcal{M}$ where $\delta:Q\times A_k\to Q$ (resp.
    $\tau:Q\to A$) is the transition function (resp. output function)
    of $\mathcal{M}$, such that $\tau(\delta(q_0,\rep_S(n))=w_n$ for
    all $n\ge 0$.
\end{definition}

\begin{example}\label{exa:poly}
    Again the set of squares $\mathcal{Q}$ is $S$-recognizable for the abstract
    numeration system $S=(a^*b^*\cup a^*c^*,\{a,b,c\},a<b<c)$. Indeed,
    we have
$$a^*b^*\cup a^*c^*=\varepsilon,a,b,c,aa,ab,ac,bb,cc,aaa,\ldots$$
and one can check that $\rep_S(\mathcal{Q})=a^*$ because the growth function of the language is $\#((a^*b^*\cup a^*c^*)\cap \{a,b,c\}^n)=2n+1$.
\end{example}

Theorem~\ref{the:cob2} can be generalized as follows  
\cite{Rigo&Maes:2002} or \cite[Ch.~3]{CANT}.

\begin{theorem}\label{the:cob3}
    An infinite word $w=w_0w_1w_2\cdots$ over an alphabet $A$ is
    morphic if and only if there exists an abstract numeration
    system $S$ such that $w$ is $S$-automatic.
\end{theorem}

Note that for generalization of Theorems~\ref{the:cob2} and
\ref{the:cob3} to a multidimensional setting, see Salon's
work \cite{Salon:1986,Salon:1987} and \cite{Charlier&Karki&Rigo} respectively.
Moreover, thanks to the above result, Durand's work can also to some
extent be expressed in terms of abstract numeration systems. Observe
that in Example~\ref{exa:poly} the abstract numeration system is based
on a regular language having a polynomial growth. This corresponds to
the case where the dominating eigenvalue of the matrix associated with
the morphism is $1$. Such a situation (polynomial versus exponential growth) is considered in
\cite{Durand&Rigo:2009}. Indeed, note that in the discussion about a
morphic version of Durand--Cobham's theorem in Remark~\ref{rem:coo}
we only considered morphisms with exponential growth, i.e., the
dominating eigenvalue being $>1$.

\section{Transcendental Numbers}\label{sec:trans}

This short section is based on a lecture given by B.~Adamczewski
during the last CANT summer school in Li{\`e}ge \cite[Chap.~8]{CANT}
and on \cite{Waldschmidt:2009}. We also refer the reader to
\cite{Adamczewski&Bell:2010}. It illustrates one of the strong links
existing between combinatorics on words and number theory. For a
survey on combinatorics on words, see for instance
\cite{Berstel&Karhumaki:2003,CKCW}.  Recall that a complex number
which is a root of a non-zero polynomial with rational (or
equivalently integer) coefficients is said to be {\em algebraic}. Otherwise, it is said to be {\em transcendental}. Since Borel's work, one thinks that base-$k$ expansion of algebraic
irrational numbers are ``complex'' and not much is known about their
properties.

With any infinite word $w=w_1w_2\cdots$ over the alphabet of digits
$A_k=\{0,\ldots,k-1\}$ is associated the real number
$\sum_{i=1}^{+\infty} w_i\, k^{-i}$ in $[0,1]$. Clearly, a real number
$\alpha$ is algebraic (over $\mathbb{Q}$) if and only if, for all
$z\in\mathbb{Z}$, $\alpha+z$ is algebraic. Indeed, if $\alpha$ is a
root of the polynomial $P(X)\in\mathbb{Q}(X)$, then $\alpha+z$ is a
root of $P(X-z)\in\mathbb{Q}(X)$.  Hence, we can restrict ourselves to
numbers in $(0,1)$.

Transcendence of a number whose binary expansion is Sturmian has
been proved in 1997 \cite{Ferenczi&Mauduit:1997}. 

\begin{example}\label{exa:fibo}
    Consider again the Fibonacci word
    $f=f_1f_2f_3\cdots=010010\cdots$. The real number
    $\sum_{i=1}^{+\infty} f_i\, 2^{-i}$ is transcendental.
\end{example}

Let
$k\in\mathbb{N}\setminus\{0,1\}$. The factor complexity of the $k$-ary
expansion $w$ of every irrational algebraic number satisfies
$${\lim\inf}_{n\to\infty} (p_w(n)-n)=+\infty.$$
The main tool is a $p$-adic version of the Thue--Siegel--Roth theorem due to
Ridout.

A combinatorial transcendence criterion obtained in
\cite{Adamczewski&Bugeaud&Luca:2004} using Schmidt's subspace theorem
\cite{schmidt:1980} is used to obtain the following result.

\begin{theorem}[Adamczewski and Bugeaud \cite{Adamczewski&Bugeaud:2007}]
    Let $k\in\mathbb{N}\setminus\{0,1\}$.  The factor complexity of
    the $k$-ary expansion $w$ of a real irrational algebraic number satisfies
$$\lim_{n\to +\infty}\frac{p_w(n)}{n}=+\infty.$$    
\end{theorem}

Let $k\ge 2$ be an integer. If $\alpha$ is a real irrational number
whose $k$-ary expansion has factor complexity in $\mathcal{O}(n)$,
then $\alpha$ is transcendental. Since, it is well-known
\cite{Cobham:1972} that automatic sequences have factor complexity
$p(n)\in\mathcal{O}(n)$, we can deduce that if a real irrational
number has an automatic $k$-ary expansion, then it is transcendental.

\begin{theorem}[Bugeaud and Evertse \cite{Bugeaud&Evertse:2008}]
    Let $k\ge 2$ be an integer and $\xi$ be a real irrational algebraic
    number with $0 <\xi < 1$. Then for any real number $\eta < 1/11$,
    the factor complexity $p(n)$ of the $k$-ary expansion of $\xi$
    satisfies
$$\lim_{n\to +\infty}\frac{p(n)}{n(\log n)^\eta}=+\infty.$$    
\end{theorem}

In \cite{Adamczewski&Rampersad:2008}, it is shown that the binary
expansion of an algebraic number contains infinitely many occurrences
of $7/3$-powers. Hence the binary expansion of an algebraic number
contains infinitely many overlaps.

\section{Combinatorial Game Theory} \label{sec:4}

Numeration systems, number theory and therefore formal language
theory also have interesting connections with combinatorial game theory. In classical textbooks like \cite{Hardy&Wright:1985,Berge}
allusion to the game of Nim is made. See \cite{nochance,fraenkel:survey}
for background on two player {\em
  combinatorial games}: no chance, no hidden information, same options for the two players who play alternatively, \ldots. In particular, in removal games, we are looking for a {\em
  winning strategy} which allows a player to consummate a win
regardless of the moves chosen by the other player. If such a strategy exists
for given initial conditions, it is therefore natural to ask about the
algorithmic complexity of the computation of the winning strategy. A first
question to answer is to determine the status $\mathcal{N}$ or $\mathcal{P}$ of a given
position \cite{fraenkel:survey}. 

A {\em $\mathcal{N}$-position}, or winning position, is a position for
which a winning strategy exists for the player who starts. A {\em
  $\mathcal{P}$-position} is a position for which all options lead to
a $\mathcal{N}$-position, and is thus winning for the second
player\footnote{In the game graph $\mathcal{G}$ where vertices are
  positions and directed edges are the allowed moves, the set of
  $\mathcal{P}$-positions is the kernel of $\mathcal{G}$: there is no
  move between any two $\mathcal{P}$-positions and from any
  $\mathcal{N}$-position, there exists a move to a
  $\mathcal{P}$-position.}. In the game of Nim played on two piles of
tokens, two players play alternatively and remove a positive number of
tokens from one of the piles. The player removing the last token win.
Otherwise stated, the first player unable to move loses (normal
condition). In \cite{Allouche&Shallit:2003}, connections between the
game of Nim (values of the Sprague-Grundy function) and the notion of
$2$-regular functions in the sense of Allouche and Shallit is observed
(finiteness of the $2$-kernel). In the famous {\em Wythoff's game}, an
extra move is allowed: removing the same positive number of tokens on
both piles.  The game of Nim can be easily generalized to $n$ piles of
tokens contrarily to Wythoff's game where extensions have been
presented but no suitable generalization is known: for the
$\mathcal{P}$-positions playing with an odd number of piles is similar
to the game of Nim and playing with an even number of piles is hard
\cite{DR,DR2,fraenkel:beat,fraenkel:flora}. In the last reference,
Wythoff's game is considered as a {\em Prime game}.  Informally, a
game whose generalization to more than one or two piles seems to be
very hard.

For instance, A.~Fraenkel makes great use of various numeration
systems to get characterizations of $\mathcal{P}$-positions \cite{fraenkel:heap}
. As an example, in
Wythoff's game, a position $(x,y)$ is a $\mathcal{P}$-position if and only if the $F$-representation $\rep_F(x)$ ends with an even number of zeroes and $\rep_F(y)=\rep_F(x)0$ is the left shift of the first component, where $F$ is the numeration basis given by the Fibonacci sequence from Example~\ref{exa:ns-fib} \cite{fraenkel:beat}.
Similarly, $(x,y)$ is a $\mathcal{P}$-position if there exists $n$ such that $(x,y)=(\lfloor n\alpha\rfloor,\lfloor n\alpha^2\rfloor)$ where $\alpha$ is the Golden ratio. So complementary Beatty sequences also enter the picture of combinatorial games \cite{DR3,fraenkel:beat,Crisp&Moran:1993}.

In \cite{DFNR}
moves that can be adjoined without changing the set of $\mathcal{P}$-positions
are characterized using the formalism of morphisms and the fact that
the computation of the successor function in the Fibonacci system is
realized by a finite transducer \cite{Frougny:1997}. Let $a\in\mathbb{N}\setminus\{0,1\}$. In the parameterized version of Wythoff's game where a player can
remove $k$ tokens from one pile and $\ell$ from the other \cite{fraenkel:beat}, with the condition $|k-\ell|<a$, the Ostrowski numeration system \cite{Berthe:2001}
based on the 
convergents of a continued fraction is used.

It is interesting to note that obviously the $\mathcal{P}$-positions of
Wythoff's game are also characterized by the Fibonacci word introduced
in Example~\ref{exa:fibo}. The $n$th $\mathcal{P}$-position is given
by the pair of indices of the $n$th symbol $0$ and $n$th symbol $1$
occurring in the Fibonacci word. This simple observation relates
combinatorial properties of morphic words like the Fibonacci or
Tribonacci words to characterizations of $\mathcal{P}$-positions of
games \cite{DR,DR2,fraenkel:flora}. Morphic characterizations of $\mathcal{P}$-positions seems to recently raise some interest among combinatorial game theorists \cite{fraenkel:flora}.

\section{Applications for Verification of Infinite State Systems}\label{sec:5}

Sets of numbers recognized by finite automata arise when analyzing
systems with unbounded mixed variables taking integer or real values.
Therefore are considered systems such as timed or hybrid automata
\cite{Boigelot&Bronne&Rassart:1997}. One needs to develop data structures representing sets manipulated during the exploration of infinite state systems. For instance, it is often needed to compute the set of reachable configurations of such a system.

Let $k\ge 2$ be an integer. Recall that A set $X\subseteq\mathbb{R}$
is {\em $k$-recognizable} if there exists a B\"uchi automaton
accepting all the $k$-representations of the elements in $X$. This
notion extends to subsets of $\mathbb{R}^d$ and to {\em Real Vector
  Automata}\index{Real Vector Automaton}\index{automaton!real vector}
or {\em RVA}\index{RVA}. Also B\"uchi--Bruy\`ere's Theorem giving a
first order logical characterization of $k$-recognizable sets of
integers holds in this context of real numbers for a suitable
structure $\langle\mathbb{R},\mathbb{Z},+,0,<,V_k\rangle$, see
\cite{Boigelot&Rassart&Wolper:1998}. Roughly speaking definability in
$\langle\mathbb{R},\mathbb{Z},+,0,<\rangle$ of subsets of
$\mathbb{R}^d$ is the natural extension of ultimately periodicity of subsets
in $\mathbb{N}$.

\begin{theorem}{\em\cite{Boigelot&Jodogne&Wolper:2005}}
    If a subset $X\subseteq\mathbb{R}^d$ is definable by a first-order
    formula in $\langle\mathbb{R},\mathbb{Z},+,0,<\rangle$, then 
     $X$ written in base $k\ge 2$ is recognizable by a weak deterministic RVA $\mathcal{A}$. 
\end{theorem}

Weakness means that each strongly connected component of $\mathcal{A}$
contains only accepting states or non-accepting states.

\begin{theorem}{\em\cite{Boigelot&Brusten:2009}}\label{the:coreal}
    Let $k,\ell\ge 2$ be two multiplicatively independent integers. If
    $X\subseteq\mathbb{R}$ is both $k$- and $\ell$-recognizable by two
    weak deterministic RVA, then it is definable in
    $\langle\mathbb{R},\mathbb{Z},+,0,<\rangle$.
\end{theorem}

The extension of Cobham--Semenov's theorem for subsets of $\mathbb{R}^d$
in this setting is discussed in \cite{Boigelot&Brusten&Leroux:2009}.
The case of two coprime bases was first considered in
\cite{Boigelot&Brusten:2009}.  Though written in a completely different language,
a similar result was independently obtained in \cite{Adamczewski&Bell:preprint}. This latter paper is
motivated by the study of some fractal sets.

\begin{remark}
    Weak deterministic RVA have a particular interest from an
    algorithmic point of view. They recognize languages that are
    recognizable by deterministic B\"uchi and deterministic co-B\"uchi
    automata. For instance, minimization algorithms in
    $\mathcal{O}(n\log n)$ exist for this class \cite{Loding}.
\end{remark}

\section{Abridged Bibliographic Notes}\label{sec:6}

With a gentle introduction to the logical formalism, a good way to
start with integer base numeration systems is to consider
\cite{Bruyere&Hansel&Michaux&Villemaire:1994}. Each time I come back
to this very well written survey, I learn something new. Then, it is a
good idea to move to the ``state of the art'' linear numeration basis
where the characteristic polynomial of the recurrence is the minimal
polynomial of a Pisot number \cite{Bruyere&Hansel:1997}. In parallel,
one should consider Frougny's chapter \cite[Chap.~7]{Lothaire:2002}
and her very interesting work on the normalization map
\cite{Frougny:1992} and beta-expansions \cite{Frougny&Solomyak:1992}.
As a good textbook on some of the aspects presented here, consider
\cite{Allouche&Shallit:2003}. The original paper of Cobham
\cite{Cobham:1972} is also worth of reading. For some general surveys
on factor complexity and the Thue--Morse word, without any required background, see
\cite{Allouche:1994,Allouche&Shallit:1999}.

Then I cannot resist advertising \cite{CANT} where in the spirit of 
Lothaire's series, we try to present the fruitful links existing
between combinatorics on words, automata theory and number theory. It
presents in a self-contained expository book much more material than is presented in this survey (ergodic theory, Rauzy fractal, joint spectral radius,\ldots).

For a list of pointers on Cobham's theorem in various contexts, see
\cite{Durand&Rigo:2010} for an updated survey. Accounts of
Perron--Frobenius theory can be found in many classical textbooks, but probably \cite{Lind&Marcus:1995} is worth reading.

Connections between symbolic dynamics and formal language theory are
fruitful: for the reader with no background in dynamics (for instance,
no knowledge in measure theory is required) and on a very introductory
level, consider \cite{Silva:2008}. Then, move to the survey \cite{Barat&Berthe&Liardet&Thuswaldner:2006} and \cite{PytheasFogg:2002}.

\section{Some Open Problems}\label{sec:7} We conclude with some general (and
probably quite hard) open problems.

\begin{itemize}
  \item As mentioned in Section~\ref{sec:sub}, the most general version
    of Cobham's theorem still relies on some mild assumptions about
    the considered morphisms (details are not given in this survey).
    F.~Durand refers to these as ``{\em good substitutions}''.  One could
    hope to relax these hypotheses and still get the same result with
    full generality \cite{Durand&Rigo:2010}. Up to now there is no
    proof of a Cobham-like theorem for a substitution having no main
    sub-substitution having the same dominating eigenvalue like
    $a\mapsto aa0$, $0\mapsto 01$ and $1\mapsto 0$. In this latter
    example, the dominating eigenvalue is $2$ but the substitution
    restricted to the alphabet $\{0,1\}$ has $(1+\sqrt{5})/2$ as
    dominating eigenvalue.
  \item Come back again to Cobham's theorem but this time for Gaussian
    integers $\mathbb{G} = \{ a+ib \mid a,b\in \mathbb {Z} \}$.
    Indeed, these numbers have nice representations using the so-called
    canonical number systems \cite{Katai&Szabo:1975}. For canonical
    numeration systems in algebraic number fields, every integer has a
    unique finite expansion which is computed starting with the least
    significant digit first. A Cobham-like conjecture for Gaussian
    integers \cite{Hansel&Safer:2003} is related to the famous Four
    Exponentials conjecture: {\em let $\{ \lambda_1 , \lambda_2\}$ and
      $\{x_1 , x_2 \}$ be two pairs of rationally independent complex
      numbers.  Then, one of the numbers $e^{\lambda_1x_1}$,
      $e^{\lambda_1 x_2}$, $e^{\lambda_2 x_1}$, $e^{\lambda_2 x_2}$ is
      transcendental}, for instance see \cite{Waldschmidt:2000}.
  \item The philosophy of Cobham's theorem also appears when considering
    {\em self-generating sets} as introduced by Kimberling \cite{Kimberling:2000}.
    For instance, consider the affine maps
    $f:\mathbb{N}\to\mathbb{N},x\mapsto 2x+1$ and
    $g:\mathbb{N}\to\mathbb{N},x\mapsto 4x+2$. A self-generating set 
    obtained from $f$ and $g$ can be defined as the smallest subset
    $S$ of $\mathbb{N}$ containing $0$ and such that $f(S)\subset S$
    and $g(S)\subset S$. In our example, the first few elements in $S$ are 
$$0, 1, 2, 3, 5, 6, 7, 10, 11, 13, 14, 15, 21, 23, 26, 27, 29, 30, 31, 42, 43, 
47, 53,\ldots.$$
One can therefore study the $k$-recognizability of
$S$. 
If one considers maps where the multiplicative constants are
multiplicatively independent, then Allouche, Shallit and Skordev
conjectured that the corresponding set cannot be $k$-recognizable \cite{Allouche&Shallit&Skordev:2005}. With
some technical hypothesis about the multiplicative coefficients when
there are at least three affine maps, this conjecture has been proved to be true 
in \cite{Karki&Lacroix&Rigo}. 
One could hope to prove this conjecture in full generality. A possible connection with smooth numbers (having only small prime factors in their decomposition) has been pointed out by J.~Shallit.

\item In combinatorial game theory the {\em Sprague-Grundy function} $g$ is
  of great interest. For instance, the positions for which $g$
  vanishes are exactly the $\mathcal{P}$-position of the game and when
  considering {\em sums of games} (several games are played simultaneously
  and at each turn, the player chooses on which of those games he will
  made a move), it can be used to distinguish
  $\mathcal{N}$-positions \cite{nochance}. For Wythoff's game, little
  is known about this function (see for instance \cite{fraenkel:sg})
  even if its recursive definition is simple. The
  value of $g(x,y)$ is the minimum excluded value (Mex) of the set of
  $g(u,v)$ where $(u,v)$ is ranging amongst the options reachable from
  $(x,y)$. By definition ${\rm Mex}\, \emptyset=0$ and ${\rm Mex}\,
  S=\min (\mathbb{N}\setminus S)$ for all finite set $S$.
$$\begin{array}{c|ccccccccccc}
 &0&1&2&3&4&5&6&7&8&9&\cdots\\
\hline
0&0&1&2&3&4&5&6&7&8&9&\cdots\\
1&1&2&0&4&5&3&7&8&6&10\\
2&2&0&1&5&3&4&8&6&7&11\\
3&3&4&5&6&2&0&1&9&10&12\\
4&4&5&3&2&7&6&9&0&1&8\\
5&5&3&4&0&6&8&10&1&2&7\\
\vdots& & & & & & & & & & & \ddots\\
\end{array}$$
    Let $F$ be the Fibonacci numeration basis. As suggested by the
    developments considered in \cite{DFNR} could the above infinite array 
     reveal some morphic structure, like having a
    finite $F$-kernel where this set could be defined as the set of
    subarrays
$$(g(x,y))_{\rep_F(x)\in\{0,1\}^*u,\ \rep_F(y)\in\{0,1\}^*v}$$
for given suffixes $u,v$? For the generalization of $k$-kernel, see for instance \cite{Rigo&Maes:2002}.

\item Theorem~\ref{the:coreal} is a Cobham-like theorem for sets of
  real numbers, definability in
  $\langle\mathbb{R},\mathbb{Z},+,0,<\rangle$ being the counterpart to
  ultimate periodicity of a set of integers. Can a simpler proof of
  this result be achieved, for instance by considering the techniques
  developed in \cite{Adamczewski&Bell:preprint}? Also could this
  result be extended to other kind of representations of real numbers.
  For instance, considering $\beta$-expansions of real numbers, we
  could define $\beta$-recognizable sets of real numbers and consider
  two multiplicatively independent real numbers $\alpha,\beta>1$. As a
  first step (and to mimic what has chronologically been done for sets
  of integers), one could consider a set of real numbers
  $X\subseteq\mathbb{R}$ which is both $k$-recognizable and $\beta$-recognizable
  by two weak deterministic RVA, with $k\ge 2$ an integer and $\beta$ a
  Pisot number like the Golden ratio, and ask is $X$ definable in
  $\langle\mathbb{R},\mathbb{Z},+,0,<\rangle$?

\item About abstract numeration systems, several questions about
  $S$-recognizable sets are open. For instance, is there some
  reasonable logical characterization of the $S$-recognizable sets of
  integers which could be compared to the characterization in the
  extended Presburger arithmetic $\langle\mathbb{N},+,V_k\rangle$. But
  one can notice that in general, if $X$ and $Y$ are
  $S$-recognizable, there is no reason to have a $S$-recognizable set
  $X+Y$ (even when considering multiplication by a constant). Another
  question is to relate the growth function $n\mapsto\#(L\cap A^n)$ of
  the regular language $L$ on which the abstract numeration system $S$
  is based and the $S$-recognizable set. For instance, if
  $P\in\mathbb{N}[X]$ is a polynomial such that $P(\mathbb{N})$ is
  $S$-recognizable, what can be said about the growth function of the
  language of numeration. Results like the one found
  in \cite{BLS} could be of interest.

\item Recently numeration systems based on the powers of a rational
  number have been introduced \cite{AFS} (motivated by a number
  theoretic question from Mahler). These numerations reveal
  interesting and intriguing properties. For instance, little is known
  about the properties of the language of numeration $L_{3/2}$. For a
  given prefix $w$, compute the number of words of length $n$ in the
  quotient $w^{-1}L_{3/2}$.

\item It is well-known since the work of Cobham \cite{Cobham:1968}
  that a morphic infinite word $w=\tau(\sigma^\omega(a))$ where
  $\sigma$ and $\tau$ are arbitrary morphisms (where both morphisms
  can be erasing and $\tau$ is not necessarily a coding) can be
  generated by a non-erasing morphism $\mu$ and a coding $\nu$. See
  for instance \cite{Allouche&Shallit:2003} for a comprehensive proof
  or \cite{Honkala:2009} for an alternative presentation. All the known
  proofs rely on morphisms and are quite long: could one describe in
  the formalism of automata theory a somehow simpler proof?

\item Let me also mention Hollander's conjecture when for a linear
  numeration basis $U$, the dominant root condition is not satisfied
  \cite{Hollander:1998}. He has conjectured that $\rep_U(\mathbb{N})$
  can be regular only if there exists $n$ such that
  $$\lim_{j\to\infty}U_{jn+k}/U_{(j-1)n+k}$$ exists and is independent
  of $k$, and the characteristic polynomial $p(X)$ of $U$ is such that
  $p(X)=q(X^n)$ where $q(X)$ is the minimal polynomial for a
  recurrence which gives a regular language \cite{Boyd:1994}.
  
\item Let $p$ be a prime. Derksen proved that the zero set of a linear
  recurrence over a field of characteristic $p$ is $p$-automatic
  \cite{Derksen,Adamczewski&Bell:2010}. Could such a result and
  Cobham's theorem be used to get back the classical
  Skolem--Mahler--Lech theorem (the zero set of a linear recurrence
  over a field of characteristic $0$ is ultimately periodic)?

\item The reader fond of logic could also look back at the list of
  open problems given by Michaux and Villemaire
  \cite{Michaux&Villemaire:open}. This survey paper is devoted to
  problems related to B{\"u}chi's characterization of sets of natural
  numbers recognizable by finite automata in base $k$, as well as to
  Cobham's and Semenov's extensions of it. 
\end{itemize}

\section*{Acknowledgments}
I would like to thank Boris Adamczewski, Val\'erie Berth\'e, V\'eronique Bruy\`ere, Eric
Duch\^ene, Narad Rampersad for the careful reading of a first draft of
this paper. They provided me useful suggestions to improve this text.

\bibliographystyle{abbrv} 
\bibliography{survey}

\end{document}